\documentstyle[twocolumn,epsf,floats,aps,prb,amssymb]{revtex}
\begin{document}
\draft
\wideabs{ 
\title{
Quasiparticle Band Structure and Density Functional Theory: \\ 
Single-Particle Excitations and Band Gaps in Lattice Models \\
}
\author{D.W. Hess}
\address{
Center for Computational Materials Science \\
Naval Research Laboratory \\
Washington, D.C. \ 20375-5345
}
\author{J.W. Serene }
\address{ Department of Physics \\
Georgetown University \\
Washington, D.C. \ \ 20057
}
\vspace{0.25in}
 \date{\today}
\vspace{0.25in}
\maketitle
\begin{abstract}
 We compare the quasiparticle band structure for a model insulator
 obtained from the fluctuation exchange approximation (FEA) with
 the eigenvalues of the corresponding density functional theory 
 (DFT) and local density approximation (LDA).  The discontinuity in 
 the exchange-correlation potential {\em for this model} is small 
 and the FEA and DFT band structures are in good agreement.  In 
 contrast to conventional wisdom, the LDA for this model 
 {\em overestimates} the size of the band gap.
 We argue that this is a consequence of an FEA self-energy that is 
 strongly frequency dependent, but essentially local. 

\end{abstract}
}

\section{Introduction}
\narrowtext

 Hohenberg and Kohn\cite{HK} showed that the total energy 
 of the interacting electron gas in an external potential is 
 given by the minimum of a functional of the electron density,
 $n({\bf r})$.
 This seminal result was followed a year later by the
 Kohn--Sham \cite{dft2} `trick' which reduces the original problem to 
 an auxiliary problem involving fictitious non-interacting particles 
 moving in a one-body effective potential composed of the external
 potential, the familiar Hartree potential, and an
 exchange-correlation potential, $v_{xc}[n]$, which is a universal 
 nonlocal functional of the spatially inhomogeneous density. This 
 innovation, together with local approximations for
 the exchange-correlation potential (LDA), underlies electronic structure 
 calculations for a wide variety of bulk materials, surfaces, and
 superlattices. Properties that can be obtained from total energy 
 calculations are generally found 
 to be accurate within a few percent.\cite{reviews}   

 A by-product of the Kohn-Sham (KS) approach is a set of single-particle
 eigenvalues. Without rigorous basis (even for the exact density
 functional theory), these are often taken to be single-particle excitation 
 energies.  This interpretation, together with the LDA, leads to reasonably 
 good agreement with experimentally determined single-particle energies 
 for many materials,\cite{reviews} but encounters difficulties in
 numerous other materials ranging from commonplace semiconductors to
 exotic f-electron intermetallics.  Here we report on calculations that 
 explore the relation between quasiparticle energies, eigenvalues obtained 
 from density functional theory (DFT), and those from a corresponding
 LDA.

 Notable among those materials for which LDA eigenvalues do not provide a correct
 account of single-particle excitation energies are seemingly ordinary 
 insulators such as Si and diamond for which the LDA eigenvalues lead to sizeable 
 {\em underestimates} of band gaps.  A more dramatic discrepancy occurs for the 
  strongly-correlated heavy-electron materials. For this class of 
 rare-earth and actinide compounds, effective masses obtained from 
 electronic structure calculations in the LDA differ by one to two 
 orders of magnitude from those observed in de Haas van Alphen experiments.\cite{HRS}  
 Among the ground states exhibited by the heavy-electron materials are
 insulators characterized by small gaps to low-energy charge excitations, 
 which display experimental signatures of heavy-electron metals upon doping. 
 In contrast to `ordinary' insulators, LDA electronic structure calculations 
 seem to {\em overestimate} the size of energy gaps in 
 these `Kondo' insulators as observed\cite{kondoi} for Ce$_3$Bi$_4$Pt$_3$.
 The insulating parent compounds of 
 the high-temperature superconductors are also 
 examples of materials for which the LDA does not adequately describe electronic 
 correlations; the LDA erroneously predicts that these are metals.\cite{HighTcref}  
 In the absence of a rigorous theoretical connection between the eigenvalues of 
 density functional theory and single-particle excitation energies, it has become a 
 matter of debate whether improved approximation to the exchange-correlation potential can remove 
 discrepancies between calculated eigenvalues and observed single-particle excitation 
 energies.

 The only rigorous connection\cite{shamkohn} between KS 
 eigenvalues and quasiparticle energies is
 that the eigenvalue of the highest occupied KS wave function is equal 
 to the chemical potential, $\mu$.  This result may be used 
 to calculate the band gap of an insulator,\cite{ss,pl} 
 \begin{equation}
  E_g = E_{cbm}^{N+1} - E_{vbm}^{N}, 
 \end{equation}
 where $E_{cbm}^{N+1}$ is the KS eigenvalue of the appropriate conduction band
 minimum calculated for the $N+1$-particle system and $E_{vbm}^{N}$ is the
 KS eigenvalue of the appropriate valence band maximum for the (insulating) 
 $N$-particle system.  Here, any ambiguity in $\mu$ for a semiconductor 
 is removed by taking appropriate limits for an $M$ particle system as 
 $M \rightarrow N+1$ from above and $M \rightarrow N$ from below.
 The ``band gap problem'' with the KS eigenvalues\cite{ss,pl} appears
 if the gap is expressed soley 
 in terms of the eigenvalues of the $N$-particle system $E^{N}$, 
 \begin{equation}
 E_g = E_{cbm}^{N} - E_{vbm}^{N} + \Delta V_{xc},
 \end{equation}
 where $\Delta V_{xc}$ is a spatially constant discontinuity of the
 exchange-correlation potential in going from an $N$ particle system
 with an exactly filled valence band to an $N+1$ particle system with
 exactly one electron in the conduction band.  
 Because (for a macroscopic
 solid) the electron densities in these two cases differ by ${\cal O}
 (1/N)$, $\Delta V_{xc}$ is in a sense maximally nonlocal: it depends
 on the total electron count (integrated over the entire system)
 and on the specific periodic structure of the solid. 

 Insofar as $\Delta V_{xc}$ is small for a particular insulator, the 
 major source of error in a LDA calculation 
 of its gap lies in the approximation for the remaining part
 of $V_{xc}$, which is taken to be local, possibly with gradient 
 corrections.\cite{perdew}  Otherwise, $\Delta V_{xc}$ 
 must be explicitly included in an $N$ particle LDA calculation for the gap.  

 There may be no universal solution to these problems; $\Delta V_{xc}$ may 
 be large for one class of systems and small for others, and the physics
 underlying the contribution of $\Delta V_{xc}$ may also differ from
 case to case.  To address this issue, one needs quasiparticle energies,
 true DFT eigenvalues, and LDA eigenvalues for the same system.  Unfortunately,
 one or another of these are known exactly in only a few special cases, and
 there is no even remotely realistic case for which all three are known
 exactly.  Hence, one is always comparing approximate results, and a 
 major concern is whether these comparisons are meaningful. 

  An essentially exact LDA is known for 3D continuum systems, but in this
  case neither the true DFT eigenvalues nor the exact quasiparticle energies
  are known with any reliability.   For this case, Godby, Schl\"uter, 
  and Sham \cite{godby} studied semiconductors and tried to obtain 
  meaningful comparisons by working consistently within the GW approximation, 
  which was used to calculate the LDA exchange-correlation potential, 
  quasiparticle energies, and the true exchange-correlation potential for 
  Si, GaAs, and AlAs.  They found that the discontinuity in the 
  exchange-correlation potential accounted for much 
  of the underestimate of the band gap in the LDA.

 Other groups have worked with 1D model Hamiltonians with a small number of
 lattice sites, for which essentially exact quantum Monte Carlo calculations
 are possible.\cite{gunn86,knorr92}  Problems with this approach include the
 relevance of results from 1D (with its well-known pathologies)
 to more realistic systems,\cite{nonfermi} the lack of accurate
 quasiparticle energies for comparison, and the question of what constitutes
 an appropriate LDA for comparison with the exact results.\cite{criticalsham}
 These references find that the discontinuity in the exchange-correlation 
 potential is small,\cite{gunn86} or large.\cite{knorr92} 

 In the spirit of Godby, Schl\"uter and Sham,\cite{godby} we have 
 studied the DFT, LDA,
 and quasiparticle energies for a 2D two-band model insulator on a bipartite 
 lattice with Hubbard on-site interactions.  In place of the GW approximation
 we use the fluctuation exchange approximation\cite{bsw} (FEA), 
 a fully self-consistent
 conserving approximation,\cite{baym} to calculate quasiparticle 
 band structures and the
 thermodynamic potential.  Because we work at finite temperatures, 
 we use Mermin's finite-temperature generalization\cite{mermin} of
 the Hohenberg-Kohn theorems to provide the connection between true DFT and the
 FEA: we define our ``full DFT'' to reproduce the site densities of the FEA
 and we define our LDA for the exchange-correlation potential from the 
 FEA for the single-band Hubbard model, which plays a role analogous to that 
 of the homogeneous interacting electron gas in continuum DFT.   We consider 
 model parameters for which no fluctuation channel is strongly dominant, but for 
 which spin fluctuations and fluctuations in the particle-particle channel provide 
 larger contributions to the thermodynamic potential than do density fluctuations.

 Calculations for our model insulator show that the FEA band gap 
 is slightly smaller than that for the DFT and that the band gap for the 
 LDA is significantly {\em larger} than 
 the FEA gap.  These relationships among gaps differ significantly from 
 the conventional electronic structure wisdom. We believe that the 
 relationships we find are a consequence of a strongly frequency-dependent but 
 essentially local self-energy.  Our self-energies are calculated
 using a propagator-renormalized theory in which the bare parameters of the
 Hamiltonian have been eliminated from thermodynamic potential.\cite{dedominicis}
 This propagator-functional formalism casts the thermodyanmic potential 
 as a functional of observable spectral functions.\cite{massrenorm} From this 
 viewpoint, the form of the self-energy is a spectral property with generalizable 
 physical consequences that transcend the specific underlying Hamiltonian. The 
 ubiquitous quasiparticles of Landau's Fermi liquid theory provide a 
 familiar example. They arise not only in liquid $^3$He 
 and in metals, but appear also in self-consistent calculations based on Hubbard 
 Hamiltonians in three dimensions.  In this spirit, we belive that the effects 
 of the locality and non-locality of the self-energy transcend the underlying 
 Hamiltonians from which they were deduced, and so the essential physics
 distilled from our results may be applied to the `Kondo insulators,' 
 such as Ce$_3$Bi$_4$Pt$_3$, for which electronic 
 structure calculations in the LDA {\em overestimate} band gaps.\cite{kondoi} 

 In the next section, we introduce single-band and bipartite-lattice
 Hubbard models and the fluctuation exchange approximation. 
 In Section III the construction of a suitable density 
 functional theory and local density approximation is described. 
 In Section IV, the LDA and FEA are compared.  Our conclusions are 
 presented in Section V.

 \section{The Model Insulator}

To explore the relation of the eigenvalues of density functional
theory to single-particle excitation energies, we take a Hubbard model
on a bipartite lattice for our model insulator.  The lattice consists of
two interpenetrating square 
 (Bravais) lattices, `A' and `B;' an A-site's 
 nearest neighbors are B-sites and its next nearest neighbors 
 are A-sites.  
The Hamiltonian for our model is
\begin{eqnarray}
   H  = \; &- \; & \sum_{(i,j) \ \sigma} t_{ij} ( c^{\dagger}_{i\sigma}c_{j\sigma}
                 + c^{\dagger}_{j\sigma}c_{i\sigma} )  \nonumber \\
	&& \; \; \; + \sum_{i \sigma} v_i n_{i \sigma}   
           + U \sum_{i} n_{i\uparrow} n_{i\downarrow} ,
           \label{hubbardh}
\end{eqnarray}
 where $c_{j\sigma} (c^{\dagger}_{j\sigma})$ annihilates (creates) a 
 particle on site $j$, $t_{ij}$ are hopping matrix elements, $v_i$ is an
 on-site single particle potential that takes the value $v_A$ ($v_B$)
 for a site on the A (B) sublattice, and
 $U$ is the intra-site repulsion experienced by two electrons on the 
 same lattice site.  We work in 2D and
 include nearest-neighbor $t$ and next-nearest-neighbor $t_{xy}$ 
 hopping amplitudes to avoid known pathologies 
 that result from Fermi surface nesting and van Hove 
 singularities at the Fermi surface. 

 We write the one-body 
 potential for the A (B) sites in the suggestive form
 $v_{A} =  v_0 - \Delta_0$ ($v_{B} = v_0 + \Delta_0 $). 
 This potential doubles the unit cell leading to a two-dimensional 
 analog of the Su, Schrieffer, and Heeger Hamiltonian for 
 polyacetylene.\cite{su79}   
 For $U=0$, the single-particle excitation energies are 
\begin{equation}
E_{\bf k}^{\pm} = \gamma_{\bf k} + 
  v_0 - \mu \pm \sqrt{|\alpha_{\bf k}|^2 + \Delta_0^2},
\label{ekfree}
\end{equation}
where
\begin{eqnarray}
\gamma_{\bf k} & = & -2 \; t_{xy} \; ( \cos k_x + \cos k_y ), \nonumber \\
\alpha_{\bf k} & = & -4 \; t \; \cos k_x \cos k_y  e^{i (k_x + k_y)/2}.
\label{gamma}
\end{eqnarray}

For nonzero $U$, the exact solution of this model is unknown.  We
use the fluctuation exchange approximation (FEA), 
based on the propagator renormalized perturbation theory of Luttinger 
and Ward, to provide quasiparticle energies and the basis for
constructing full and LDA density functional theories.  

 The Luttinger and Ward \cite{lw} formula expresses the grand thermodynamic 
 potential $\Omega$ as a functional of the fully renormalized Green's 
 function $G$ and the self energy $\Sigma$,  
 \begin{equation}
  \Omega (T, \mu) = -2 \; {\rm Tr} \ [ \Sigma G + \ln
 (- \; G_0^{-1} + \Sigma)] + \Phi [G].  \label{feaomega}
 \end{equation}
 Here `Tr' indicates a sum over momentum, frequency, and 
 sublattice arguments,\cite{note1} and $G_0$ is the Green's function 
 for the noninteracting system ($U=0$) which contains the bare 
 band structure. When viewed as a 
functional of $G$ and $\Sigma$,
$\Omega$ is stationary with respect to independent
variations around the true $G$ and $\Sigma$ for a given $G_0$ and $U$, 
 leading to the `skeleton diagram' expansion
for the self-energy, 
\begin{equation}
\Sigma({\bf k}, \varepsilon_n)   =   \frac{1}{2} \ \frac{\delta \Phi [G]}
                        {\delta G({\bf k}, \varepsilon_n)},  \label{delphi} 
\end{equation}
and to Dyson's equation, 
\begin{equation}
G^{-1}({\bf k}, \varepsilon_n)   =   G^{-1}_0({\bf k}, \varepsilon_n)
                  - \Sigma({\bf k}, \varepsilon_n) \; .       \label{dyson}
\end{equation}
The self-consistent solutions of Eqs. (\ref{delphi}) and (\ref{dyson}) 
completely specify $G$ and $\Sigma$, and together with Eq. (\ref{feaomega}) 
provide a self-consistent description of single-particle excitation spectra and
thermodynamic properties.  The FEA entails a specific prescription for
$\Phi[G]$ that includes contributions from the exchange of
spin, density, and Cooper-pair fluctuations in addition to 
self-consistent contributions at first (Hartree Fock) and second 
order in the interaction.  The propagator for our model is 
a $2\times 2$ matrix labeled by sublattice indices; 
a more detailed discussion of the FEA 
for the Hubbard model on a bipartite lattice, together with explicit 
expressions for $\Phi[G]$, appears in the Appendix.

We have solved Eqs. (\ref{delphi}) and (\ref{dyson}) self-consistently 
on a $64 \times 64$ mesh using a parallel algorithm described 
elsewhere.\cite{shmb7}  This is not our most accurate method
for solving the equations of the FEA. 
The more accurate algorithm of Ref. \onlinecite{deisz95}, which does 
not contain a high-frequency cutoff in the traditional sense, is
essential for calculating temperature derivatives of the thermodynamic
potential and for performing calculations near an instability to 
an ordered state.  Since none of these conditions apply here, 
the algorithm of Ref. \onlinecite{shmb7} suffices for our purposes.

The A- and B-site densities needed for the 
density functional theory calculations described below are 
obtained directly from the diagonal components of the propagator,
\begin{equation}
n_i  = {2 {T}\over{N}}\sum_{{\bf k},n}  G_{ii}({\bf k}, \varepsilon_n),  
\label{nanb}
\end{equation}
where $T$ is the temperature and $N$ is the number of ${\bf k}$-points.
Because the FEA is a conserving approximation, this is equivalent to 
calculating the site densities from the thermodynamic potential,
\begin{equation}
n_i = {{\partial \; \Omega[T, v_A, v_B]} \over {\partial \; v_i}}.
\end{equation}
Explicit calculations show that this internal self-consistency is 
achieved to better than $ 1 \%$.

 \section{Density Functional Theory for Lattice Models}

  Following Mermin,\cite{mermin} it is straightforward to show that 
  the thermodynamic potential of our model is a stationary function 
  of the A- and B-site densities. 
  The thermodynamic potential (with arguments $T$ and $\mu$ suppressed) is 
  \begin{equation}
  \Omega [\{n_j\}] = \Theta [\{n_j\}] + \Omega_{H}[\{n_j\}] + 
  \Omega_{xc}[\{n_j\}] + \Omega_{ext}[\{n_j\}], 
  \label{dft}
  \end{equation}
  where $\Omega_{ext}$ is the contribution from the external
  potential, $\Theta + \Omega_{ext}$ is the grand thermodynamic 
  potential of a system of fictitious non-interacting particles with 
  site densities $\{n_j\}$, and $\Omega_{H}$, and $\Omega_{xc}$ 
  are the Hartree and  exchange-correlation potential contributions. 
  For our model, the Hartree contribution to $\Omega$ is
  \begin{equation}
  \Omega_H [n_A, n_B] = \frac{1}{4} \; (n_A^2 + n_B^2) \; U,
  \end{equation}
  and the contribution from the external potential $v_{j}$ is 
 \begin{equation}
  \Omega_{ext}[\{n_j\}] = \sum_j v_{j} n_j.
 \end{equation}
  The condition that the grand thermodynamic potential be stationary
  with respect to variations in the density leads to the set of coupled
  equations,
  \begin{equation} 
 \frac{ d \Theta[ \{ n_j \} ] }{d n_i} + 
 \frac{ d \Omega_{ext}[ \{ n_j \}]}{d n_i} +
 \frac{ d \Omega_{H}[ \{ n_j \} ]}{d n_i} + 
 \frac{ d \Omega_{xc}[ \{ n_j \} ]}{d n_i}  = 0 . \label{ddft}
  \end{equation}

  \subsection{ Full Density Functional Theory}
  
   The KS formulation at finite temperature invokes
   an auxiliary system of noninteracting particles 
   in a one-body effective potential $V_{\rm eff}$, that is
   defined by writing Eq. (\ref{ddft}) in the form,
  \begin{equation} 
    \frac{ d \Theta[ \{ n_j \} ] }{d n_i} + V_{\rm eff}[\{n_j\}] = 0 \; .
   \label{vareqn}
  \end{equation} 
   The thermodynamic potential for the auxiliary problem is
   \begin{equation}
   \Omega_{\rm eff}^0 = -2 T \sum_{\alpha} 
   \ln \left[ 1 + e^{-\beta(\epsilon_\alpha - \mu)} \right], 
   \end{equation}
   where the single-particle eigenvalues $\epsilon_\alpha$ 
   satisfy the Schr\"{o}dinger equation 
   \begin{equation}
   (K + V_{\rm eff}[\{ n_j \}]) \; \psi_{\alpha} = 
    \epsilon_{\alpha} \psi_{\alpha} \; ,    \label{kseqn}
   \end{equation}
   and $K$ is the hopping Hamiltonian (first term of Eq. (\ref{hubbardh})).
    The site densities at finite temperature are related to the 
   KS wave functions through 
   \begin{equation}
    n_i = 2 \sum_{\alpha}\; | \psi_{\alpha}({\bf r}_i) |^2 f(\epsilon_{\alpha}),
    \label{ksdens}
   \end{equation}
    where $f(\epsilon)$ is the usual Fermi function. Because 
    Eq. (\ref{vareqn}) is also the KS variational condition for the auxiliary
    problem with $V_{eff}$ regarded as an external potential, densities  
    that result from the self-consistent solution of 
    Eqs. (\ref{kseqn}) and (\ref{ksdens}) are guaranteed to be identical
    to those of the original interacting system. 

    For our model, $V_{xc} [ \{ n_j \} ] $ is unknown; we use the 
    FEA site densities together with the KS formulation to find
    the exchange correlation potential, 
    $V_{xc} = (v^A_{xc}, v^B_{xc})$.  To do this, we find 
    one body potentials $v^A_{\rm eff}$ and $v^B_{\rm eff}$, for which
    the site-densities of the finite-temperature auxiliary KS problem 
    are the {\em same} as the FEA 
    site-densities, and invert the definition of $V_{\rm eff}$ 
    to obtain $V_{xc}$. 
    A convenient expression (equivalent to Eq. (\ref{ksdens}))
    for the site-densities of the non-interacting problem is obtained by 
    taking the trace of the site-diagonal components of the propagator 
    for the non-interacting system given in  Eq. (\ref{freegreen}).   
    The full DFT band structure can be extracted from Eqs. (\ref{ekfree}) 
    using $v_i = v^i_{\rm eff}$.

    Because the FEA is a conserving approximation, the FEA grand thermodynamic 
    potential has the same stationary properties with respect to variations of $G$
    and $\Sigma$ as the exact grand thermodynamic potential.  As a consequence, the
    exchange-correlation potential that we have calculated is identical to that
    calculated from Sham's integral equation\cite{lusham} for $V_{xc}$.

  \subsection{ The Local Density Approximation}

  For our model, the single-band Hubbard model plays the role 
  analogous to that of the uniform interacting electron gas in 
  the formulation of the continuum LDA for electronic structure 
  theory. So, we use Eq. (\ref{dft}) to calculate $\Omega_{xc}[n]$ in 
  the FEA for the single-band Hubbard model 
  (the Hamiltonian of Eq. \ref{hubbardh} with $v_i = 0$) with 
  the same ``Coulomb potential,'' $U$.  To avoid possible confusion, we 
  emphasize that in this case the electron density is the same at every site, 
  and $\Omega_{xc}$ becomes a simple function of the uniform density, $n$.
  This entails evaluating Eq. (\ref{feaomega}) with the
  self-consistently calculated FEA propagator for the single-band Hubbard model,
  evaluating the Hartree contribution to the thermodynamic potential,
  and rearranging Eq. (\ref{dft}) with $\Omega_{ext} =0$
  to extract $\Omega_{xc}$.
  The LDA exchange-correlation potential is 
  \begin{equation}
   V_{xc}^{LDA}(n) = \frac{d \Omega_{xc}(n)}{d n}. \label{vxclda}
  \end{equation}
  In this way the LDA exchange-correlation potential corresponding to 
  the full DFT is calculated in an approximation equivalent to that 
  of the FEA calculation. 

  Given $V_{xc}^{LDA}(n)$, Eqs. (\ref{kseqn}) and (\ref{ksdens}) can 
  be solved directly with external potentials $v_0$ and $\Delta_0$,
  subject to the constraint that the total density agrees with that of
  the FEA calculation.  As in the case of the DFT, Eqs. (\ref{ekfree}) 
  with the potentials $v^{LDA}_{i, \; {\rm eff}}$ determine the LDA
  band structure. 

 \section{Results}

   We focus on single-band and two-band models with next-nearest-neighbor 
   hopping matrix element $t_{xy} = 0.35$ and $U = 3$.  This set of parameters 
   avoids possible complications of dominant 
   fluctuation contributions in any one specific channel as might arise as a 
   consequence of, for example, Fermi surface nesting.\cite{note2}  
   For the model insulator (see below), the contributions to the grand thermodynamic 
   potential from the Hartree and second-order diagrams are the largest. Somewhat 
   smaller spin, density, and Cooper pair fluctuation contributions are all roughly 
   the same order. For $T = 0.02$, explicit values for the $\Phi$-functionals in their 
   respective channels are:
   $\Phi_{2} = -1.87 \times 10^{-2}$, $\Phi_{\rm pp} = 5.06 \times 10^{-3}$,
   $\Phi^{df}_{ph} = 8.41 \times 10^{-4}$, and 
   $\Phi^{sf}_{ph} = -3.31 \times 10^{-3}$. 

\begin{figure}[h]
 \vspace{.1in}
  \epsfxsize=3.300in\centerline{\epsffile{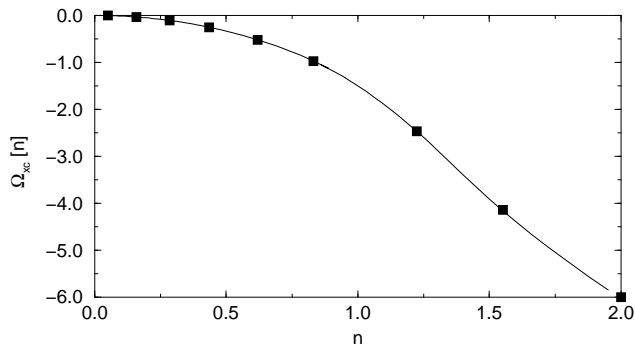}}
 \vspace{.15in}
  \caption{
  The exchange-correlation contribution to the thermodynamic
  potential $\Omega_{xc}$ as a function of density for the single-band
  Hubbard model in the fluctuation exchange approximation (FEA) for
  $U=3$ at two temperatures.  The solid line is a fit to 77 points
  (see text) for $T = 0.08$. The widely spaced points are calculated
  for $T = 0.02$.}
  \label{f:omxc}
  \end{figure}

\begin{figure}[h]
  \vspace{.1in}
  \epsfxsize=3.300in\centerline{\epsffile{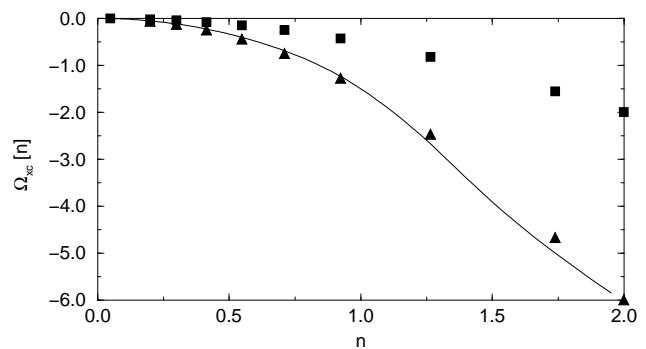}}
  \vspace{.15in}
  \caption{The exchange-correlation contribution to the thermodynamic
   potential $\Omega_{xc}(n)$  for $U = 3 $ (solid line; see previous
   figure and text), $U=1$ ($\blacksquare$), and the $U=1$ result scaled
   by a factor of 3 ($\blacktriangle$).
   }
\label{omxctwou}
  \end{figure}

  The exchange-correlation contribution to the thermodynamic
  potential $\Omega_{xc}$ as a function of density for a
  single-band Hubbard model with next-nearest-neighbor 
  hopping matrix element $t_{xy} = 0.35$, $U = 3$ and $T = 0.08$ 
  is shown in Fig. \ref{f:omxc} (we henceforth measure energies in 
  units of the nearest-neighbor hopping matrix element $t$).  
  The solid line is a polynomial fit of the form,
  \begin{equation}
    \Omega_{xc}(n) =  \sum_{i=2}^{10}  a_i n^i \label{omxc}
  \end{equation}
   through 77 data points.  The coefficients, presented
   in Table 1, lead to a fit with an error of less than $1 \%$ over the 
  entire density range; the worst agreement occurring at low density.
\begin{table}[h]
\caption{Values of the coefficients $a_i$ in Eq. (\ref{omxc})
for the exchange-correlation contribution to the grand thermodynamic
potential as a function of density. These parameters are 
for $U =3$, $T=0.08$, and $t_{xy}=0.35$.} 
\begin{tabular}{ccc}
$a_2$ = \dec -1.32532  & $a_3$ = \dec -0.300919  & $a_4$ = \dec   3.27475 \\ 
$a_5$ = \dec -13.7277  & $a_6$ = \dec 29.0402    & $a_7$ = \dec -34.1589  \\ 
$a_8$ = \dec  21.9011  & $a_9$ = \dec -7.11752   & $a_{10}$ = \dec   0.917821  \\
\end{tabular}
\end{table}
  It is interesting to note, in contrast to the 
  exchange-correlation energy used in by Ref. \onlinecite{gunn86},
  that there is no evidence for a significant $n^{4/3}$ contribution to 
  $\Omega_{xc}$.\cite{note2} 
  Fig. \ref{f:omxc} also shows $\Omega_{xc}(n)$ for $T = 0.02$ at several 
  densities, from which we conclude that $\Omega_{xc}(n)$ is 
  weakly temperature dependent over the range $0.02 < T < 0.08$.
  The function $\Omega_{xc}(n)$ is to a reasonable approximation 
  linear in $U$ (see Fig. \ref{omxctwou}); 
  the largest deviations occur at high density as might be expected.
  We calculate the local density approximation 
  of the exchange correlation potential from Eq. (\ref{omxc}). 

\begin{figure}[h]
  \vspace{.1in}
  \epsfxsize=3.400in\centerline{\epsffile{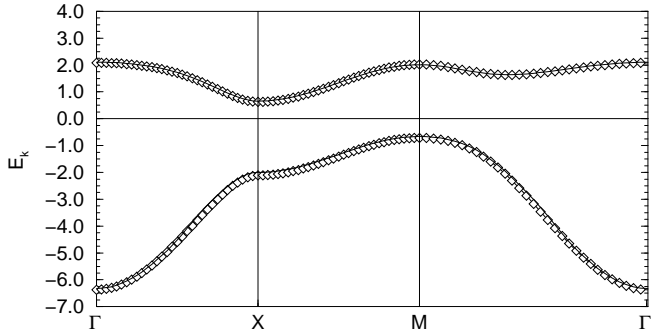}}
  \vspace{.15in}
  \caption{
   The band structure for ${\bf k}$ along high symmetry directions
   in the Brillouin zone from the FEA calculation (solid) compared
   with that from the DFT ($\diamond$) for $T=0.08$, $U=3$, and 
   $t_{xy} = 0.35$. }
   \label{bsfeadft}
  \end{figure}

\begin{figure}[h]
  \vspace{.1in}
  \epsfxsize=3.300in\centerline{\epsffile{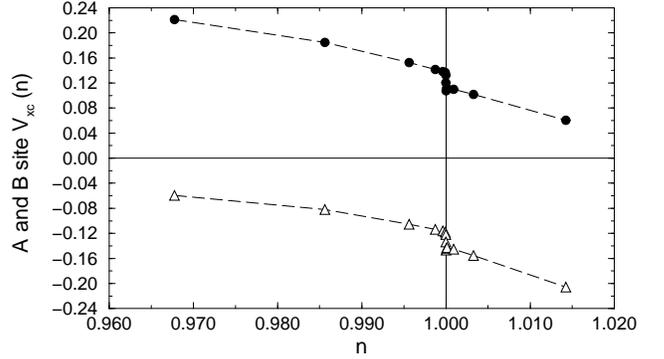}}
  \vspace{.15in}
  \caption{
   The `exact' exchange-correlation potential
   $V_{xc}$ (see text) for A-($\bullet$) and B-sites($\triangle$). The
   discontinuity is $\sim 0.02$ and occurs at half filling.}
   \label{xctvxc}
  \end{figure}

  We now compare the FEA single-particle excitations of an insulator 
  to the eigenvalues obtained from the full DFT and those from the LDA.
  To this end we calculate the fully self-consistent self-energy in 
  the FEA for a bare\cite{note2a} staggered potential 
  $\Delta_0 = 2.532$. The quasiparticle band structure is 
  obtained from the zeros of the real part of the denominator
  of the retarded Green's function $G^{\rm R}$.  The fully renormalized 
  retarded propagator was constructed using Dyson's equation and the 
  retarded self-energy. The latter was obtained from an analytic 
  continuation of the self-energy from the imaginary-frequency axis
  to the real-frequency axis using N-point Pad\'{e} Approximants.\cite{vs}  
  The resulting band structure for
  $\Delta_0 =2.532$ and $n=1.000$ (with $n_A^{FEA} = 1.688$ and 
  $n_B^{FEA} = 0.312$), shown in Fig. \ref{bsfeadft} for ${\bf k}$ along the 
  high-symmetry path $\Gamma - X  - M - \Gamma$ in the square zone, 
  is that of an insulator with an indirect gap 
  $E_g = E_X^+ - E_M^- = 1.318$.  The corresponding full DFT
  band structure is found by evaluating the (bare) dispersion 
  relations in Eq. (\ref{ekfree})
  using on-site one body potentials $v, \Delta$ for which the site 
  densities $n_A$ and $n_b$ are equal to $n_A^{FEA}$ and $n_B^{FEA}$.
  Starting from
  the (renormalized) on-site single-body potentials of the FEA
  $(v^{FEA}, \Delta^{FEA})$, a Newton-Raphson method allowed us to 
  determine (to 1 part in $10^8$) $v^{DFT} = -0.745$ and 
  $\Delta^{DFT} = 1.368$.
   The DFT band structure appears in comparison with that for 
   the FEA in Fig. \ref{bsfeadft}.  The fact that these band structures are
   essentially indistinguishable suggests that the discontinuity in 
   the exchange-correlation potential is small.  In Fig. \ref{xctvxc}
   we show the `exact' $V_{xc}^{FEA}(n)$ obtained from $v^{DFT}$
   and $\Delta^{DFT}$ for a range of densities about 
   $n = 1$ and observe that, as expected, $V_{xc}^A$ and $V_{xc}^B$ each
   show a small discontinuity $|\Delta V_{xc}| = 0.026$. 
    Note that the small structure appearing at the 
   high-density side ($n = 1.00^+$) of the discontinuity is not physically
   significant, but is rather a numerical artifact reflecting a finite 
   high-frequency cutoff in the FEA calculation; this feature is 
   systematically reduced as the number of Matsubara frequencies kept 
   in the calculation is increased.\cite{note3} 

\begin{figure}[h]
  \vspace{.1in}
  \epsfxsize=3.400in\centerline{\epsffile{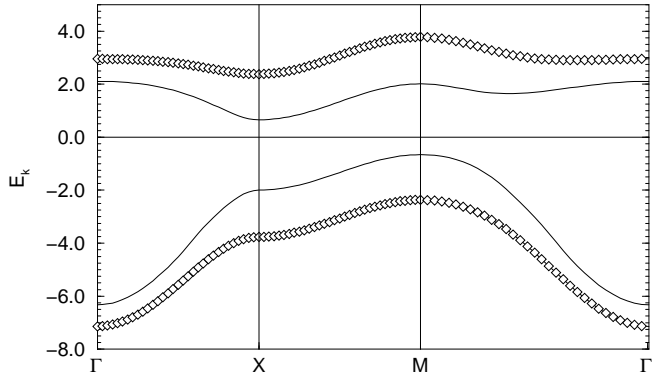}}
  \vspace{.15in}
  \caption{ A comparison of the FEA band structure (solid) with
  that of the LDA ($\diamond$), where the LDA is adjusted to yield the
  same bare staggered field and total density of the FEA
  calculation. }
\label{bsfealda}
  \end{figure}

   To calculate the band structure in the LDA, we take $V_{xc}^{LDA}$ 
   from Eqs. (\ref{vxclda}) and (\ref{omxc}),
   and solve Eqs. (\ref{kseqn}) and (\ref{ksdens}) using a 
   Newton-Raphson method to find 
   on-site one-body potentials $(v^{LDA}, \Delta^{LDA})$ that
   yield the same total density with the same bare staggered 
   potential $\Delta_0 $ as in the FEA calculation.  The LDA band
   structure obtained from Eqs. (\ref{ekfree}) 
   with $(v^{LDA}, \Delta^{LDA})$ is shown compared
   with that for the FEA in Fig. \ref{bsfealda}.   For $n \approx 1$, 
   $\partial n / \partial \mu$ is vanishingly small and a wide
   range of $v^{LDA}$ lead to a $\Delta_0$ and $n$ equal to those 
   of the FEA calculation. 
   On the other hand, $\Delta^{LDA}$ is determined to high 
   accuracy by our procedure and is largely insensitive to 
   changes in $v^{LDA}$ for $n \sim 1$.  Since $T$ is much 
   smaller than the (indirect) band gap resulting from 
   $\Delta^{LDA}$, we take $v^{LDA}$ so that the chemical 
   potential lies in the middle of the gap as expected on physical 
   grounds.  We have explicitly verified that this choice 
   leads to $n = n^{FEA}$ and $\Delta_0 = \Delta_0^{FEA}$
   to within the accuracy to which these quantities are known.

   The LDA and FEA band structures differ significantly; the LDA band 
   structure {\em cannot} be corrected by uniformly and independently 
   shifting the bands by a constant amount. The LDA band structure 
   cannot be corrected by using the ``scissors operator.''
   An examination of the spatial dependence of the FEA self-energy
   shows that the $\Sigma_{ij}$ are essentially local. 
   The self-energy is essentially uniform in $\bf{k}$ and strongly 
   frequency dependent.  The excitation energies $E_k$ are the poles of 
   the retarded propagator, $G^R$.   By combining the appropriate analytic 
   continuation of Eq. (\ref{freegreen})
   with Dyson's equation, we express $G^R$ in terms of the components 
   of the self-energy and find that the poles of $G^R$ satisfy
    \begin{eqnarray}
 \left[ E_{\bf k}  + \Delta_0 + \mu + \gamma_{\bf k} - 
  \Sigma_{AA}(E_{\bf k}) \right] \times \nonumber \\
 \; \; \; \left[ E_{\bf k} + \Delta_0 + \mu + \gamma_{\bf k} - 
  \Sigma_{BB}(E_{\bf k}) \right]   -  \nonumber \\
    \; \; \; \left[ \alpha_{\bf k} + \Sigma_{AB}(E_{\bf k}) \right]
     \left[ \alpha^{*}_{\bf k} + \Sigma_{BA}(E_{\bf k})
     \right]
     & = & 0.  \label{excitations} 
    \end{eqnarray}
    Here $\alpha_{\bf k}$ and $\gamma_{\bf k}$ are defined as in Eq.
    (\ref{gamma}) and we have used 
    $\Sigma_{ij}({\bf k},\varepsilon) \sim \Sigma_{ij}(\varepsilon)$. 
    The $E_{\bf k}$ that solves Eq. (\ref{excitations}) has a ${\bf
    k}$-dependence that not only reflects of that of $\gamma_{\bf k}$ 
    and $\alpha_{\bf k}$ but also includes implicit additional 
    ${\bf k}$-dependence through the frequency dependence of the 
    $\Sigma_{ij}$.  

    As is plainly evident from Fig. \ref{bsfealda}, 
    the LDA {\em overestimates} the the band gap by about a factor of 3. 
    In the light of LDA calculations for insulators, the result 
    $E_g^{LDA} > E_g^{FEA}$ is perhaps surprising. Since the FEA 
    and the self-consistent GW approximation can be viewed as
    propagator-functional theories, physical properties may be viewed as a 
    consequence of the form of spectral functions\cite{dedominicis} 
    rather than of the form of a particular Hamiltonian. Considered 
    in this way, results of continuum calculations are relevant insofar as
    they lead to trends in self-energies and hence spectral functions.  From
    this perspective, we believe that the explanation for our results 
    lies in the work of Godby, Schl\"{u}ter, and 
    Sham.\cite{godby}  They observed that, for the GW approximation, 
    the nonlocality of the self-energy tends to a widen the gap while 
    frequency dependence tends to narrow the gap.  Given the importance
    of the frequency dependence of $\Sigma_{ij}$ in determining the 
    FEA band structure and given that the DFT band structure is obtained 
    from dispersion relations for a non-interacting system, albeit with
    largely `renormalized' on-site potentials, the observed
    close agreement between these band structures may be of a rather subtle 
    physical origin.  This agreement may also be in part a reflection 
    of a moderate interaction strength ($U \sim 3/8$ of the bare, 
    $\Delta_0 = 0$, bandwidth) and may degrade with larger $U$. 
    It is also evident from the DFT calculation (see Fig. \ref{xctvxc}) 
    that  $V_{xc}^A$ and $V_{xc}^B$ are of opposite sign for the 
    density range in which we are working. On the other hand, 
    $\Omega_{xc}$ of Fig. \ref{f:omxc} leads to a $V_{xc}^{LDA} < 0$ over the 
    entire density range. To compensate, the LDA must increase the size 
    of $\Delta^{LDA}$, which increases the gap.

 \section{Conclusions}
 
 We have calculated the renormalized band structure and site densities
 for a model insulator in the fluctuation exchange approximation (FEA) 
 and in corresponding full and approximate (LDA) density functional theories.  
 The band structures calculated from the full density-functional theory 
 are in good agreement with the  `true' band structures obtained from 
 the poles of the retarded FEA propagator.   The gap obtained from the 
 DFT is slightly smaller than that of the FEA; the difference is accounted 
 for by a discontinuity in $V_{xc}$ that is small in magnitude. It is also 
 of opposite sign to that observed in GW calculations for `real' ordinary 
 insulators.   We find that gap in the LDA is {\em larger} than that of 
 the full DFT.  The relationship among the gaps differs from that observed 
 in continuum GW calculations for ordinary insulators.
 An examination of the FEA self-energy shows it to be essentially local, 
 suggesting that this is a consequence of the frequency dependence of the 
 self-energy. We speculate that such a relationship ought to be observed
 in the `Kondo' insulators where it is expected that, owing to strong 
 electronic correlations, the frequency dependence of the self-energy 
 is more important than its spatial nonlocality.

\acknowledgments

DWH acknowledges the support of the Office of Naval Research.
This work was supported in part by a grant of computer time from the
DoD HPC Shared Resource Center Naval Research Laboratory Connection 
Machine facility CM-5e. We would like to thank J.Q. Broughton
whose `distressing lack of humility' has been a source of inspiration 
over the years, and to wish him luck in his new endeavours.

\appendix
\section*{The FEA for a Hubbard Model on a 
Bipartite Lattice}

The single-particle propagator is central to the calculation 
of the density and single-particle excitation spectra.
It is convenient to adopt a matrix notation for the
propagator in imaginary time for the Hamiltonian in Eq. (\ref{hubbardh}), 
\begin{equation}
{\cal G}_{ij} ({\bf r}_i - {\bf r}^{\prime}_j,\tau) = -
< {\cal T} \{ \psi({\bf r}_i, \tau) \psi^\dagger({\bf r}_j^{\prime},0) \}> \; ,
\end{equation}
where ${\cal T}$ is the usual Wick's time ordering operator and 
$\psi({\bf r}_i)$ is an annihilation operator acting at
${\bf r}_i$, where $i$ is a sublattice index ($i=A,B$).  
Because of the translational invariance of the full Bravais lattice,  
the propagator is a function of a coordinate difference. 
Unlike ${\cal G}_{AA}$,
${\cal G}_{AB}$ is defined on a Bravais lattice that does not include
the origin. We find it convenient for numerical calculations on 
parallel computers to introduce a propagator $G_{ij}$ defined so that
$G_{ii} = {\cal G}_{ii}$, 
$G_{AB}({\bf r}, \tau) = {\cal G}_{AB}({\bf r} - \hat{c}, \tau)$, and
$G_{BA}({\bf r}, \tau) = {\cal G}_{BA}({\bf r} + \hat{c}, \tau)$,
where $\hat{c}$ is a vector that joins a point 
on the A sublattice to one on the B sublattice.
By this definition, the arguments of all components of the propagator range 
over the same (sub)lattice and the usual Fourier transform connects
(${\bf r},\tau$) space to (${\bf k}$, $\varepsilon_n$) for all components 
of $G_{ij}$.  

Dyson's equation (Eq. (\ref{dyson})) should now be interpreted as an 
equation for the ($2 \times 2$) matrices $G$, $G_0$ and $\Sigma$.  
Keeping only nearest-neighbor and next-nearest-neighbor hopping amplitudes, 
the propagator for the noninteracting system in (${\bf k}$, $\varepsilon_n$)
space is,
\begin{eqnarray}
\lefteqn{G_0^{-1}({\bf k}, \varepsilon_n) =} & \hfill \nonumber \\
& \; \; \; \; \; \; \; \; \left( \begin{array}{cc}
   i \varepsilon_n - \bar{\gamma}_{\bf k} + \Delta_0  & - \alpha_{\bf k} \\
          - \alpha_{\bf k}^*  & i \varepsilon_n - \bar{\gamma}_{\bf k} - \Delta_0  \\
	  \end{array}
	   \right) \; , \label{freegreen} 
\end{eqnarray}
where $\bar{\gamma}_{\bf k} = \gamma_{\bf k} + v - \mu$, and $\gamma_{\bf k}$ and $\alpha_{\bf k}$ defined as in Eq. (\ref{gamma}) 
above.

The FEA\cite{bsw} takes $\Phi [\{G_{ij}\}]$ 
to be the sum of the Hartree-Fock diagram, the single second-order diagram, 
and particle-hole and particle--particle bubble chains 
that describe exchanged density, spin-density, 
and (singlet) pair fluctuations. For the bipartite lattice, explicit 
expressions for $\Phi[\{G_{ij}\}]$ are
\begin{eqnarray}
\Phi_2 \; \, & = &  \; - \case 1/2 {\rm Tr}\left[(\chi^{ph}_{AA})^2 + 
			       (\chi^{ph}_{BB})^2 
			       + 2 \chi^{ph}_{AB} \chi^{ph}_{BA} \right]  \\  
\Phi_{ph}^{df} & = &  \; \case 1/4 {\rm Tr}\{
 	   \ln ( (1 + \chi^{ph}_{AA})(1 + \chi^{ph}_{BB}) 
	     - \chi^{ph}_{AB} \chi^{ph}_{BA} ) \nonumber  \\
		&&	     - \chi^{ph}_{AA} - \chi^{ph}_{BB} \nonumber  \\
 &&  + \case 1/2 \left[ (\chi^{ph}_{AA})^2 + (\chi^{ph}_{BB})^2 \right]
     + \chi^{ph}_{AB} \chi^{ph}_{BA} \} \\
\Phi_{ph}^{sf} & = &  \; \case 3/4 {\rm Tr} \{
			   \ln ( (1 - \chi^{ph}_{AA})(1 - \chi^{ph}_{BB}) 
		     - \chi^{ph}_{AB} \chi^{ph}_{BA} ) \nonumber \\ 
	 && 	    + \chi^{ph}_{AA} + \chi^{ph}_{BB}   \nonumber \\
 &&  + \case 1/2 \left[ (\chi^{ph}_{AA})^2 + (\chi^{ph}_{BB})^2 \right]
     + \chi^{ph}_{AB} \chi^{ph}_{BA} \} \\
\Phi_{pp} & = & \;  \case 1/4 {\rm Tr}\{
			   \ln ( (1 + \chi^{pp}_{AA})(1 + \chi^{pp}_{BB}) 
		     - \chi^{pp}_{AB} \chi^{pp}_{BA} ) \nonumber \\ 
	 &&	     - \chi^{pp}_{AA} - \chi^{pp}_{BB}  \nonumber \\
  && + \case 1/2 \left[ (\chi^{pp}_{AA})^2 + (\chi^{pp}_{BB})^2 \right]
     + \chi^{pp}_{AB} \chi^{pp}_{BA} \}. 
\end{eqnarray}
Here $\chi^{ph}_{ij}$ and $\chi^{pp}_{ij}$ are particle-hole and 
particle-particle susceptibility bubbles connecting lattice sites 
$i$ and $j$ and are related to the fully renormalized Green's function by
\begin{eqnarray}
\chi^{pp}_{ij}(&{\bf q},& \omega_m) =  U(T/N) \nonumber \\
&&  \sum_{\bf k} \sum_{n} 
   G_{ij}({\bf k}+{\bf q},\varepsilon_n +\omega_m)G_{ji}(-{\bf k},-\varepsilon_n) \\
\chi^{ph}_{ij}(& {\bf q},& \omega_m) =  - U(T/N) \nonumber \\
&&\sum_{\bf k} \sum_{n} 
  G_{ij}({\bf k}+{\bf q},\varepsilon_n +\omega_m)G_{ji}({\bf k},\varepsilon_n).
\end{eqnarray} 
The components of the self-energy matrix are obtained from Eq. (\ref{delphi}),
\begin{equation}
\Sigma_{ij}({\bf k}, \varepsilon_n)   =   \frac{1}{2} \ 
\frac{\delta \Phi [G]} {\delta G_{ji} ({\bf k}, \varepsilon_n)}. 
\label{mdelphi} 
\end{equation}

\newpage


\begin{references}
\bibitem{massrenorm} Our calculations here are based on an explicit
     realization of this systematic program at the single-particle level
     (mass renormalization).
\bibitem{note1} Here we consider only paramagnetic systems.
\bibitem{note2a} Here we use `bare' to indicate values $v_0, \Delta_0$
           that appear (or are trivially related to those that appear)
           in the Hamiltonian, in contrast to the renormalized single
           body potentials $v^{FEA}, \Delta^{FEA}$ into which the
           Hartree contribution to the self-energy has been included.
\bibitem{note3} We used 256 Matsubara frequencies to calculate 
    $\Omega_{xc}$ for the one-band Hubbard model and have explicitly 
    verified that our LDA is insensitive to any further increases in 
    the high-frequency cutoff.
\end{references}
\end{document}